\RequirePackage{amsthm}

\documentclass[pdflatex,sn-mathphys-num]{sn-jnl}

\usepackage{graphicx}%
\usepackage{multirow}%
\usepackage{amsmath,amssymb,amsfonts}%
\usepackage{amsthm}%
\usepackage{mathrsfs}%
\usepackage[title]{appendix}%
\usepackage{xcolor}%
\usepackage{textcomp}%
\usepackage{booktabs}%
\usepackage{algorithm}%
\usepackage{algorithmicx}%
\usepackage{algpseudocode}%
\usepackage{listings}%

\def\arcsec{\hbox{$^{\prime\prime}$}}
\def\approxlt{\ifmmode \rlap{$<$}{}_{{}_{{}_{\textstyle\sim}}} \else%
$\rlap{$<$}{}_{{}_{{}_{\textstyle\sim}}}$\fi}

\makeatletter
\let\jnl@style=\rm
\def\ref@jnl#1{{\jnl@style#1}}

\def\aj{\ref@jnl{AJ}}                   
\def\actaa{\ref@jnl{Acta Astron.}}      
\def\araa{\ref@jnl{ARA\&A}}             
\def\apj{\ref@jnl{ApJ}}                 
\def\apjl{\ref@jnl{ApJ}}                
\def\apjs{\ref@jnl{ApJS}}               
\def\ao{\ref@jnl{Appl.~Opt.}}           
\def\apss{\ref@jnl{Ap\&SS}}             
\def\aap{\ref@jnl{A\&A}}                
\def\aapr{\ref@jnl{A\&A~Rev.}}          
\def\aaps{\ref@jnl{A\&AS}}              
\def\azh{\ref@jnl{AZh}}                 
\def\baas{\ref@jnl{BAAS}}               
\def\bac{\ref@jnl{Bull. astr. Inst. Czechosl.}}
\def\caa{\ref@jnl{Chinese Astron. Astrophys.}}
\def\cjaa{\ref@jnl{Chinese J. Astron. Astrophys.}}
\def\icarus{\ref@jnl{Icarus}}           
\def\jcap{\ref@jnl{J. Cosmology Astropart. Phys.}}
\def\jrasc{\ref@jnl{JRASC}}             
\def\memras{\ref@jnl{MmRAS}}            
\def\mnras{\ref@jnl{MNRAS}}             
\def\na{\ref@jnl{New A}}                
\def\nar{\ref@jnl{New A Rev.}}          
\def\pra{\ref@jnl{Phys.~Rev.~A}}        
\def\prb{\ref@jnl{Phys.~Rev.~B}}        
\def\prc{\ref@jnl{Phys.~Rev.~C}}        
\def\prd{\ref@jnl{Phys.~Rev.~D}}        
\def\pre{\ref@jnl{Phys.~Rev.~E}}        
\def\prl{\ref@jnl{Phys.~Rev.~Lett.}}    
\def\pasa{\ref@jnl{PASA}}               
\def\pasp{\ref@jnl{PASP}}               
\def\pasj{\ref@jnl{PASJ}}               
\def\rmxaa{\ref@jnl{Rev. Mexicana Astron. Astrofis.}}%
\def\qjras{\ref@jnl{QJRAS}}             
\def\skytel{\ref@jnl{S\&T}}             
\def\solphys{\ref@jnl{Sol.~Phys.}}      
\def\sovast{\ref@jnl{Soviet~Ast.}}      
\def\ssr{\ref@jnl{Space~Sci.~Rev.}}     
\def\zap{\ref@jnl{ZAp}}                 
\def\nat{\ref@jnl{Nature}}              
\def\iaucirc{\ref@jnl{IAU~Circ.}}       
\def\aplett{\ref@jnl{Astrophys.~Lett.}} 
\def\apspr{\ref@jnl{Astrophys.~Space~Phys.~Res.}}
\def\bain{\ref@jnl{Bull.~Astron.~Inst.~Netherlands}} 
\def\fcp{\ref@jnl{Fund.~Cosmic~Phys.}}  
\def\gca{\ref@jnl{Geochim.~Cosmochim.~Acta}}   
\def\grl{\ref@jnl{Geophys.~Res.~Lett.}} 
\def\jcp{\ref@jnl{J.~Chem.~Phys.}}      
\def\jgr{\ref@jnl{J.~Geophys.~Res.}}    
\def\jqsrt{\ref@jnl{J.~Quant.~Spec.~Radiat.~Transf.}}
\def\memsai{\ref@jnl{Mem.~Soc.~Astron.~Italiana}}
\def\nphysa{\ref@jnl{Nucl.~Phys.~A}}   
\def\physrep{\ref@jnl{Phys.~Rep.}}   
\def\physscr{\ref@jnl{Phys.~Scr}}   
\def\planss{\ref@jnl{Planet.~Space~Sci.}}   
\def\procspie{\ref@jnl{Proc.~SPIE}}   
\makeatother

\theoremstyle{thmstyleone}%
\theoremstyle{thmstyletwo}%

\theoremstyle{thmstylethree}%

\begin{document}


\title[Article Title]{The fast X-ray transient EP\,240315a: a $z \sim 5$ gamma-ray burst in a Lyman continuum leaking galaxy}


\author[1,2]{\fnm{Andrew J.} \sur{Levan}}\email{a.levan@astro.ru.nl}

\affil[1]{\orgdiv{Department of Astrophysics/IMAPP}, \orgname{Radboud University Nijmegen}, \orgaddress{\street{P.O.~Box 9010}, \city{Nijmegen}, \postcode{6500~GL},   \country{The Netherlands}}}

\affil[2]{\orgdiv{Department of Physics}, \orgname{University of Warwick}, \orgaddress{\street{Coventry, CV4 7AL}, \country{UK}}}

\author[1]{\fnm{Peter G.} \sur{Jonker}}


\author[3]{\fnm{Andrea} \sur{Saccardi}}

\affil[3]{\orgdiv{GEPI}, \orgname{Observatoire de Paris, Université PSL, CNRS}, \orgaddress{\street{5 Place Jules Janssen}, \city{Meudon}, \postcode{92190}, \country{France}}}

\author[4,5,1]{\fnm{Daniele Bj\o{}rn} \sur{Malesani}}
\affil[4]{\orgdiv{Cosmic Dawn Center (DAWN)}, \orgaddress{\country{Denmark}}}
\affil[5]{\orgdiv{Niels Bohr Institute}, \orgname{University of Copenhagen}, \orgaddress{\street{Jagtvej 128}, \city{Copenhagen}, \postcode{2200}, \country{Denmark}}}
\author[6]{\fnm{Nial R.} \sur{Tanvir}}  
\affil[6]{\orgdiv{School of Physics and Astronomy}, \orgname{University of Leicester}, \orgaddress{\street{University Road}, \city{Leicester}, \postcode{LE1 7RH},   \country{United Kingdom}}}

\author[7,5]{\fnm{Luca} \sur{Izzo}}

\affil[7]{\orgdiv{Osservatorio Astronomico di Capodimonte}, \orgname{INAF}, \orgaddress{\street{Salita Moiariello 16}, \city{Napoli}, \postcode{80131},   \country{Italy}}}

\author[4,5,8]{\fnm{Kasper E.} \sur{Heintz}}  
\affil[8]{\orgdiv{Department of Astronomy}, \orgname{University of Geneva}, \orgaddress{\street{Chemin Pegasi 51}, \city{Versoix}, \postcode{1290}, \country{Switzerland}}}

\author[9,10]{\fnm{Daniel} \sur{Mata S\'anchez}}  
\affil[9]{\orgdiv{Instituto de Astrof\'{i}sica de Canarias}, \orgname{IAC}, \orgaddress{\street{E-38205}, \city{La Laguna, S/C de Tenerife}, \country{Spain}}}
\affil[10]{\orgdiv{Departamento de Astrof\'isica, Univ. de La Laguna},   \orgaddress{\street{E-38206}, \city{La Laguna, Tenerife},     \country{Spain}}}

\author[1]{\fnm{Jonathan} \sur{Quirola-V\'asquez}}


\author[9,10]{\fnm{Manuel} \sur{A. P. Torres}}

\author[3]{\fnm{Susanna D.} \sur{Vergani}}

\author[11]{\fnm{Steve} \sur{Schulze}}
\affil[11]{\orgdiv{Center for Interdisciplinary Exploration and Research in Astrophysics}, \orgname{Northwestern University}, \orgaddress{\street{1800 Sherman Ave, 8th Floor}, \city{Evanston}, \postcode{60201}, \state{IL}, \country{USA}}}


\author[12]{\fnm{Andrea} \sur{Rossi}}

\affil[12]{\orgdiv{Osservatorio di Astrofisica e Scienza dello Spazio}, \orgname{INAF}, \orgaddress{\street{Via Piero Gobetti 93/3}, \city{Bologna}, \postcode{40129},   \country{Italy}}}

\author[13]{\fnm{Paolo} \sur{D'Avanzo}}  
\affil[13]{\orgdiv{Osservatorio astronomico di Brera}, \orgname{INAF}, \orgaddress{\street{Via E. Bianchi 46}, \city{Merate}, \postcode{23807},   \country{Italy}}}

\author[14,15]{\fnm{Benjamin} \sur{Gompertz}}

\affil[14]{\orgdiv{School of Physics and Astronomy}, \orgname{University of Birmingham}, \orgaddress{  \city{Birmingham}, \postcode{B15 2TT}, \country{UK}}}
\affil[15]{\orgdiv{Institute for Gravitational Wave Astronomy}, \orgname{University of Birmingham}, \orgaddress{\city{Birmingham}, \postcode{B15 2TT}, \country{UK}}}

\author[16]{\fnm{Antonio} \sur{Martin-Carrillo}}

\affil[16]{\orgdiv{School of Physics and Centre for Space Research}, \orgname{University College Dublin}, \orgaddress{\street{Belfield}, \city{Dublin 4},     \country{Ireland}}}

\author[17,18]{\fnm{Antonio} \sur{de Ugarte Postigo}}

\affil[17]{\orgdiv{Universit\'{e} de la C\^ote d'Azur, Observatoire de la C\^ote d'Azur, CNRS, Artemis},   \orgaddress{  \city{Nice}, \postcode{F-06304},   \country{France}}}
\affil[18]{\orgdiv{Aix Marseille Univ, CNRS, LAM}, \orgaddress{\city{Marseille},     \country{France}}}

\author[19]{\fnm{Benjamin} \sur{Schneider}}

\affil[19]{\orgdiv{Kavli Institute for Astrophysics and Space Research}, \orgname{Massachusetts Institute of Technology}, \orgaddress{\street{77 Massachusetts Avenue}, \city{Cambridge}, \postcode{02139}, \state{MA}, \country{USA}}}

\author[20,21]{\fnm{Weimin}\sur{Yuan}}

\affil[20]{\orgdiv{National Astronomical Observatories}, \orgname{Chinese Academy of Sciences}, \orgaddress{\city{Beijing}, \postcode{100103},   \country{China}}}

\affil[21]{\orgdiv{University of Chinese Academy of Sciences}, \orgname{Chinese Academy of Sciences}, \orgaddress{\city{Beijing}, \postcode{100049},   \country{China}}
}

\author[20,21]{\fnm{Zhixing}\sur{Ling}}


\author[20]{\fnm{Wenjie}\sur{Zhang}}


\author[20,21]{\fnm{Xuan}\sur{Mao}}


\author[20]{\fnm{Yuan}\sur{Liu}}


\author[20]{\fnm{Hui}\sur{Sun}}


\author[22]{\fnm{Dong} \sur{Xu}}

\affil[22]{\orgdiv{Key Laboratory of Space Astronomy and Technology, National Astronomical Observatories}, \orgname{Chinese Academy of Sciences}, \orgaddress{\street{20A Datun Road}, \city{Beijing}, \postcode{100101},   \country{China}}}

\author[22]{\fnm{Zipei} \sur{Zhu}}


\author[23]{\fnm{Jos\'e Feliciano} \sur{Ag\"u\'i Fern\'andez}}
\affil[23]{ \orgname{Centro Astron\'omico Hispano en Andaluc\'ia}, \orgaddress{\street{Observatorio de Calar Alto, Sierra de los Filabres}, \city{G\'ergal, Almer\'ia}, \postcode{04550},   \country{Spain}}}

\author[12]{\fnm{Lorenzo} \sur{Amati}}


\author[24,25,26]{\fnm{Franz E.} \sur{Bauer}}

\affil[24]{\orgdiv{Instituto de Astrof{\'{\i}}sica, Facultad de F{\'{i}}sica and Centro de Astroingenier{\'{\i}}a, Facultad de F{\'{i}}sica, Pontificia Universidad Cat{\'{o}}lica de Chile}, \orgname{UC}, \orgaddress{\street{Campus San Joaquín, Av. Vicuña Mackenna 4860, Macul}, \city{Santiago}, \postcode{7820436}, \state{RM}, \country{Chile}}}
\affil[25]{\orgdiv{Millennium Institute of Astrophysic}, \orgname{MAS}, \orgaddress{\street{Nuncio Monse{\~{n}}or S{\'{o}}tero Sanz 100, Of 104, Providencia}, \city{Santiago}, \postcode{7500011}, \state{RM}, \country{Chile}}}
\affil[26]{\orgdiv{Space Science Institute}, \orgname{SSI}, \orgaddress{\street{4750 Walnut Street, Suite 205}, \city{Boulder}, \postcode{80301}, \state{CO}, \country{USA}}}

\author[13]{\fnm{Sergio} \sur{Campana}}


\author[27]{\fnm{Francesco} \sur{Carotenuto}}

\affil[27]{\orgdiv{Astrophysics, Department of Physics}, \orgname{University of Oxford}, \orgaddress{\street{Keble Road}, \city{Oxford}, \postcode{OX1 3RH},   \country{United Kingdom}}}

\author[28,1]{\fnm{Ashley} \sur{Chrimes}}

\affil[28]{\orgdiv{European Space Agency (ESA)}, \orgname{European Space Research and Technology Centre (ESTEC)}, \orgaddress{\street{Keplerlaan 1}, \city{Noordwijk}, \postcode{2201~AZ},   \country{The Netherlands}}}

\author[1]{\fnm{Joyce N. D.} \sur{van Dalen}}


\author[29,30]{\fnm{Valerio} \sur{D'Elia}}

\affil[29]{\orgdiv{Space Science Data Centre}, \orgname{Italian Space Agency}, \orgaddress{\street{via del Politecnico snc}, \city{Rome}, \postcode{00133},   \country{Italy}}}
\affil[30]{\orgdiv{INAF}, \orgname{Osservatorio Astronomico di Roma}, \orgaddress{\street{via di Frascati 33}, \city{Monteporzio Catone (RM)}, \postcode{00078},   \country{Italy}}}

\author[7]{\fnm{Massimo} \sur{Della Valle}}

\author[31]{\fnm{Massimiliano} \sur{De Pasquale}}

\affil[31]{\orgdiv{MIFT Department}, \orgname{University of Messina}, \orgaddress{\street{Via F. S. D'Alcontres 31}, \city{Messina}, \postcode{98166},   \country{Italy}}}

\author[32,9]{\fnm{Vikram S.} \sur{Dhillon}}

\affil[32]{\orgdiv{Department of Physics and Astronomy}, \orgname{University of Sheffield}, \orgaddress{  \city{Sheffield}, \postcode{S3 7RH},   \country{United Kingdom}}}

\author[33,34]{\fnm{Llu\'is} \sur{Galbany}}

\affil[33]{\orgdiv{Institute of Space Sciences (ICE-CSIC)},   \orgaddress{\street{Campus UAB, Carrer de Can Magrans, s/n}, \city{Barcelona}, \postcode{E-08193},   \country{Spain}}}
\affil[34]{\orgdiv{Institut d'Estudis Espacials de Catalunya (IEEC)},   \orgaddress{\city{Castelldefels}, \postcode{E-08860},   \country{Spain}}}

\author[1]{\fnm{Nicola} \sur{Gaspari}}

\author[35]{\fnm{Giulia} \sur{Gianfagna}}

\affil[35]{\orgdiv{Istituto di Astrofisica e Planetologia Spaziali}, \orgname{INAF}, \orgaddress{\street{via Fosso del Cavaliere 100}, \city{Rome}, \postcode{I-00133},   \country{Italy}}}

\author[36]{\fnm{Andreja} \sur{Gomboc}}
\affil[36]{\orgdiv{Center for Astrophysics and Cosmology}, \orgname{University of Nova Gorica}, \orgaddress{\street{Vipavska cesta 11 c}, \city{Ajdov\v s\v cina}, \postcode{SI-5270},   \country{Slovenia}}}

\author[6]{\fnm{Nusrin} \sur{Habeeb}}


\author[1]{\fnm{Agnes P. C.} \sur{van Hoof}}


\author[13]{\fnm{Youdong} \sur{Hu}}


\author[37]{\fnm{Pall} \sur{Jakobsson}}

\affil[37]{\orgdiv{Centre for Astrophysics and Cosmology}, \orgname{University of Iceland}, \orgaddress{\street{Dunhagi}, \city{Reykjav\'ik}, \postcode{107},   \country{Iceland}}}

\author[6]{\fnm{Yashaswi} \sur{Julakanti}}


\author[38]{\fnm{Judith} \sur{Korth}}
\affil[38]{\orgdiv{Lund Observatory}, \orgname{Lund University}, \orgaddress{\street{Box 118}, \city{Lund}, \postcode{22100},   \country{Sweden}}}

\author[39]{\fnm{Chryssa} \sur{Kouveliotou}}
\affil[39]{\orgdiv{Astrophysics Office/ZP12, NASA Marshall Space Flight Center}, \orgname{MSFC}, \orgaddress{\city{Huntsville}, \postcode{35812}, \state{Alabama}, \country{U.S.A.}}}

\author[40,1]{\fnm{Tanmoy} \sur{Laskar}}
\affil[40]{\orgdiv{Department of Physics \& Astronomy}, \orgname{University of Utah}, \orgaddress{\city{Salt Lake City}, \postcode{84112}, \state{Utah}, \country{USA}}}

\author[32]{\fnm{Stuart P.} \sur{Littlefair}}


\author[12]{\fnm{Elisabetta} \sur{Maiorano}}


\author[41]{\fnm{Jirong} \sur{Mao}}
\affil[41]{\orgdiv{Yunnan Observatories}, \orgname{Chinese Academy of Sciences}, \orgaddress{\city{Merate}, \postcode{23807}, \state{Yunnan Province}, \country{China}}}

\author[30]{\fnm{Andrea} \sur{Melandri}}


\author[42]{\fnm{M. Coleman} \sur{Miller}}

\affil[42]{\orgdiv{Department of Astronomy and Joint Space-Science Institute}, \orgname{University of Maryland}, \orgaddress{\city{College Park}, \postcode{20742}, \state{Maryland}, \country{USA}}}

\author[43,44]{\fnm{Tamal} \sur{Mukherjee}}
\affil[43]{\orgdiv{School of Mathematical and Physical Sciences}, \orgname{Macquarie University}, \orgaddress{\street{Street}, \city{Sydney}, \postcode{2109}, \state{NSW}, \country{Australia}}}
\affil[44]{\orgdiv{ARC Centre of Excellence for All Sky Astrophysics in 3 Dimensions}, \orgname{ASTRO3D}, \orgaddress{\street{Street},       \country{Australia}}}

\author[45]{\fnm{Samantha R.}\sur{Oates}}
\affil[45]{\orgdiv{Department of Physics}, \orgname{Lancaster University}, \orgaddress{\street{Bailrigg}, \city{Lancaster}, \postcode{LA1 4YW}, \state{Lancashire}, \country{UK}}}

\author[6]{\fnm{Paul} \sur{O'Brien}}


\author[3]{\fnm{Jesse T.} \sur{Palmerio}}

\author[10,9]{\fnm{Hannu} \sur{Parviainen}}


\author[1]{\fnm{Dani\"elle L. A.} \sur{Pieterse}}

\author[30]{\fnm{Silvia} \sur{Piranomonte}}

\author[35]{\fnm{Luigi} \sur{Piro}}


\author[46]{\fnm{Giovanna} \sur{Pugliese}}

\affil[46]{\orgdiv{Anton Pannekoek Institute of Astronomy}, \orgname{University of Amsterdam}, \orgaddress{\street{Science Park 904}, \city{Amsterdam}, \postcode{1098 XH},   \country{The Netherlands}}}

\author[1]{\fnm{Maria E.} \sur{Ravasio}}


\author[6]{\fnm{Ben} \sur{Rayson}}


\author[47]{\fnm{Ruben} \sur{Salvaterra}}

\affil[47]{\orgdiv{Istituto di Astrofisica Spaziale e Fisica cosmica}, \orgname{INAF}, \orgaddress{\street{Via Corti 12}, \city{Milano}, \postcode{20133},   \country{Italy}}}

\author[48]{\fnm{Rub\'en} \sur{S\'anchez-Ram\'irez}}

\affil[48]{\orgdiv{Instituto de Astrof\'isica de Andaluc\'ia}, \orgname{CSIC}, \orgaddress{\street{Glorieta de la Astronom\'ia, s/n}, \city{Granada}, \postcode{E-18080},   \country{Spain}}}

\author[49,50]{\fnm{Nikhil} \sur{Sarin}}
\affil[49]{\orgdiv{The Oskar Klein Centre, Department of Physics}, \orgname{Stockholm University}, \orgaddress{\street{AlbaNova}, \city{Stockholm}, \postcode{SE-106 91}, \state{Stockholm}, \country{Sweden}}}
\affil[50]{\orgdiv{Nordita}, \orgname{Stockholm University and KTH Royal Institute of Technology}, \orgaddress{\street{Hannes Alfvéns väg 12}, \city{Stockholm}, \postcode{SE-106 91}, \state{Stockholm}, \country{Sweden}}}

\author[45]{\fnm{Samuel P. R.} \sur{Shilling}}


\author[6]{\fnm{Rhaana L. C.} \sur{Starling}}


\author[13]{\fnm{Gianpiero} \sur{Tagliaferri}}


\author[35]{\fnm{Aishwarya Linesh} \sur{Thakur}}


\author[51]{\fnm{Christina C.} \sur{Th\"one}}

\affil[51]{\orgdiv{ASU-CAS}, \orgname{Astronomical Institute, Czech Academy of Sciences}, \orgaddress{\street{Fri\v cova 298}, \city{Ond\v rejov}, \country{Czech Republic}}}

\author[52]{\fnm{Klaas} \sur{Wiersema}}

\affil[52]{\orgdiv{Centre for Astrophysics Research}, \orgname{University of Hertfordshire}, \orgaddress{\street{College Lane}, \city{Hatfield}, \postcode{AL10 9AB},   \country{United Kingdom}}}

\author[14,15]{\fnm{Isabelle} \sur{Worssam}}


\author[43]{\fnm{Tayyaba} \sur{Zafar}}


\abstract{The nature of the minute-to-hour long Fast X-ray Transients (FXTs) localised by telescopes such as \emph{Chandra}, \emph{Swift}, and \emph{XMM-Newton} remains mysterious, with numerous models suggested for the events. Here, we report multi-wavelength observations of EP240315a, a 1600\,s long transient detected by the Einstein Probe, showing it to have a redshift of $z=4.859$.  We measure a low column density of neutral hydrogen, indicating that the event is embedded in a low-density environment, further supported by direct detection of leaking ionising Lyman-continuum. The observed properties are consistent with EP\,240315a being a long-duration $\gamma$-ray burst, and these observations support an interpretation in which a significant fraction of the FXT population are lower-luminosity examples of similar events. Such transients are detectable at high redshifts by the Einstein Probe and, in the (near) future, out to even larger distances by SVOM, THESEUS, and Athena, providing samples of events into the epoch of reionisation.}




\maketitle

\section{Introduction}\label{sec1}

Studying the nature of short-timescale astrophysical transient events is a major endeavour across the electromagnetic spectrum, from millisecond activity in fast radio bursts (FRBs \cite{petroff19}) to the multiple sources that evolve on timescales of hours to days in the optical sky \cite{abbott_bns,prentice18}, to the high-energy gamma-ray bursts (GRBs) that have been studied for more than half a century. Perhaps surprisingly, the origin of similar singular outbursts that have been detected using instruments sensitive to soft X-ray photons ($\approx0$.3--10 keV) remains uncertain. These fast X-ray transients (FXTs) have been known for almost as long as GRBs \citep[e.g.][]{cooke76,rappaport76}, however, it is only more recently that better and occasionally rapid localizations have allowed some of them to be pinpointed as securely extragalactic events \citep{soderberg08,jonker13}. 

Intensive archival searches have now found $>30$ FXTs in \emph{Chandra} \citep{jonker13,glennie15,bauer17,lin22,qv22,qv23} and \emph{XMM-Newton} \citep{alp20,novara20} observations. These transients are characterized by durations of hundreds to thousands of seconds \citep{qv22,qv23,alp20} and power-law spectra. There have been many suggestions for their origin, which include powering by rapid spin-down of a millisecond magnetar formed in a binary neutron star merger \cite{2013ApJ...763L..22Z}; by the tidal disruption (TDE) of a white dwarf by an intermediate-mass black hole \cite{jonker13,MacLeod2016};  by a supernova shock breakout \cite{Waxman2017}; or by cocoon-like emission from a jet breakout in long-GRBs \cite{2017ApJ...834...28N}. 

The relationship of these FXTs to previously identified populations of high-energy transients remains, however, unclear. In particular, while GRBs were identified based on their $E \times F_E$ spectral peak in the 100 keV -- 1 MeV range (where $F_E$ is the energy flux density and $E$ the photon energy), missions with softer responses, such as \textit{Beppo}\/SAX and HETE-2, also uncovered GRB-like events with much lower peak energies, often dubbed X-ray flashes (XRFs; \citep[e.g.][]{sakamoto04}). These have durations more like those of classical GRBs, but spectra that peak at tens of keV, or sometimes even $<10$ keV. In at least two cases, the very long (2000\,s) GRB/XRF\,060218 \cite{pian06} and GRB/XRF\,020903 \cite{soderberg04_020903} the detection of supernova signatures points to similar progenitors for XRFs as for long-duration GRBs, although other events have strong limits on supernova emission \cite{levan05,soderberg05}. However, the relation of these XRFs to the longer-lived FXTs is less certain. In this regard, it is worth noting that the list of models for GRB creation postulated before the identification of the first afterglows was very long (118 distinct models according to the list of \cite{nemiroff94}), so there is no shortage of potential routes to the creation of high-energy emission. The question of whether these scenarios manifest in nature, and at what rate, remains open. 

The recent launch of the Einstein Probe (EP; \cite{EP,Yuan2022}) offers the opportunity to make decisive inroads into these questions. In particular, its Wide-field X-ray Telescope (WXT)  has a soft X-ray response (0.5--4 keV), a wide field of view (3600 deg$^2$) and unprecedented sensitivity thanks to the use of focussing Lobster-eye optics.
It should enable the discovery of substantial populations of high-energy transients. The expectation is that the soft response and detection via imaging, rather than scintillators or coded-mask detectors should enhance the detectability of GRBs (and related phenomena) at high redshift where both time dilation and the redshift of the spectrum could favor X-ray detection methods. Discovery of distant transients would be of significant importance because of their value as probes of the high-redshift Universe \cite{tanvir21}. In this regard, GRBs and related transients have several advantages over other probes, even galaxies now identified at extreme redshifts by JWST. GRBs signal the collapse of individual massive stars at very high redshift, potentially even providing a route for the identification of first-generation stars. Moreover, because these GRBs select galaxies based on a single star rather than the integrated light, they can sample galaxies across the luminosity function, including those undetectable to current technology \cite{tanvir12}. Finally, the afterglows themselves are bright backlights on which we view the imprint of the interstellar and intergalactic medium. This enables, for example, the reconstruction of detailed abundance patterns in high redshift galaxies \cite{heintz23}, direct measurements of the neutral fraction of the intergalactic medium during the epoch of reionisation \cite{fausey24}, and a powerful route to determine the fraction of photons that can escape from these regions of massive stars to drive this reionisation \cite{tanvir19,vielfaure20}.

\section{Results}\label{sec2}
EP\,240315a was discovered by the EP WXT as a soft X-ray transient with 1600~s duration \citep{GCN35931}. Although no associated 
GRB was promptly reported, analysis of data from both \emph{Swift}-BAT and Konus-\textit{Wind} showed a 50 s long GRB with a consistent sky position which occurred $\sim 400$ seconds after the onset of emission detected by EP \cite{svinkin24,delaunay24}. In general, GRBs show an evolution from hard to soft, and so both the difference in X-ray and $\gamma$-duration and time of onset would appear, at first sight, unusual. However, it should be noted that GRB follow-up is normally initiated by the $\gamma$-ray trigger with X-ray observations occurring later. X-rays before the main $\gamma$-ray pulse have not been well studied, although at least a few examples (GRB\,011121, GRB\,981226, GRB\,980519) exist in the sample of GRBs observed by \emph{Beppo}\/SAX \cite{piro05, 2000ApJ...540..697F, 1999ApJ...516L..57I}. Overall, the duration observed in $\gamma$-rays would mark EP\,240315a as a normal long-GRB while the 1600\,s emission in the X-ray regime would imply an ultra-long GRB \cite{levan14}. The field covering EP\,240315a was imaged by the ATLAS survey approximately one hour after the outburst, and a bright new optical source, AT\,2024eju, was uncovered in these observations \cite{srivastav24,gillanders24}. 

We obtained spectroscopic observations with the GTC/OSIRIS+ and VLT/X-shooter instruments beginning 27 and 29 hours after the burst \cite{qv24, 2024GCN.35936....1S}, respectively (1 and 3 hours after the report of the counterpart). These spectra show a strong break at $\sim 7120$\,\AA{} which can be interpreted as due to Ly-$\alpha$ absorption. The X-shooter observations reveal numerous high ionisation lines from metals including carbon, nitrogen, oxygen and silicon, and from these, we infer a systemic redshift of 
$z = 4.8585 \pm 0.0001$ 
(see Figure~\ref{spec} and Methods). Our spectroscopy provides good signal-to-noise from 7100 to 15,000\,\AA{} (1216--2600 \AA{} in the rest frame). Over this range, the optical counterpart light is reasonably described by a spectral slope of $F_{\nu} \propto \nu^{-1.0}$, typical of GRB afterglows. A second X-shooter spectrum was obtained on 31 March 2024. At this epoch, the afterglow had faded substantially, but we recovered a Ly-$\alpha$ line from the host galaxy in emission. 

\begin{figure*}
\begin{center}
\centerline{
\includegraphics[angle=0,width=\textwidth]{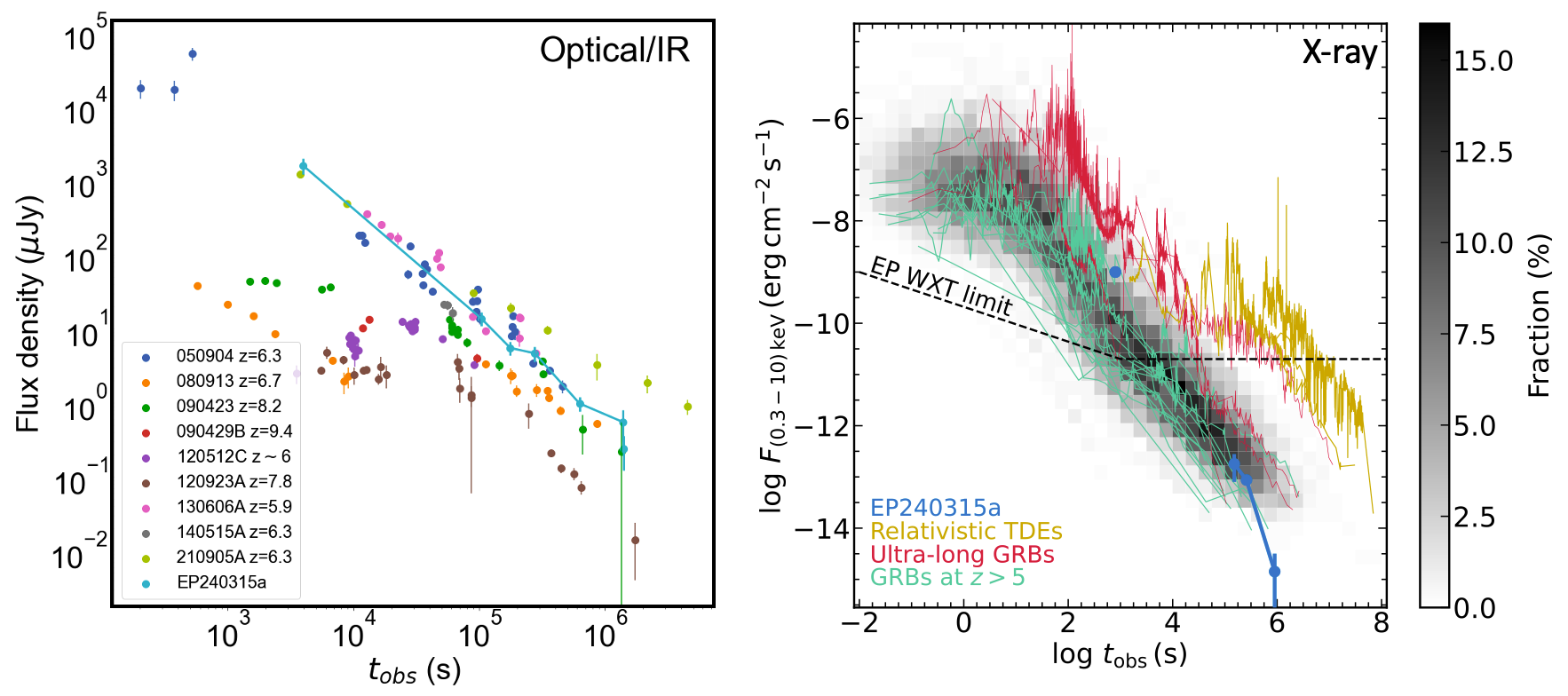}}
\end{center}
\caption{{\bf A comparison of the X-ray and optical/IR properties of EP\,240315a with a sample of high redshift and ultra-long duration GRBs}. Left: A comparison of the optical afterglow of EP\,240315a with several high-redshift GRBs. The afterglow is plotted in the $z$ band with the exception of the first point which is corrected from the ATLAS $c$ band (see \cite{srivastav24,gillanders24}) via a colour term derived by folding the X-shooter spectrum through the $c$-band response. The afterglow of EP\,240315a is a bright, but typical high-redshift afterglow. Right: The location of EP\,240315a in the X-ray flux space, compared to \emph{Swift}-GRBs, high-$z$ bursts, ultra-long GRBs and relativistic TDEs. While the observations of EP\,240315a extend to later times and fainter fluxes than many \emph{Swift} observations they lie in the same regime as normal long and ultra-long GRBs, and the long duration ($>1000$~s) above the EP WXT limit would be observed for many \emph{Swift} events.}
\label{xray_flux_lc}
\end{figure*}

In addition to these spectroscopic observations, we also obtained optical and IR photometric observations with the VLT, GTC, NOT, TNG and LBT, and X-ray observations with the \emph{Chandra} X-ray Observatory (see Figure~\ref{xray_flux_lc} and Methods). Combining these with data from the literature (see Methods) the optical/IR counterpart decays in the $z$-band as $t^{-1.4 \pm 0.2}$ from the time of our first to last observations. The X-ray observations favour a steeper slope, with the two \emph{Chandra} epochs suggesting a decay faster than $t^{-2.1}$ (at 95\% confidence). The late-time X-ray slope would appear to be indicative of a GRB afterglow post-jet break as favoured by detailed afterglow modelling (see \cite{Liu-et-al}). The difference in X-ray and optical slopes may indicate a contribution from an underlying host galaxy in our latest $z$-band data-point, although the late-time detections are only marginal. 

The combined spectral and temporal behaviour of the counterpart is consistent with that seen for the afterglows of GRBs. In Figure~\ref{xray_flux_lc} we compare the behaviour of EP\,240315a with the X-ray and optical/IR lightcurves of a sample of both high redshift GRBs detected with \emph{Swift} and with the population of ultra-long GRBs \cite{levan14}. The properties of EP\,240315a are consistent with these comparison events, although it is notable that while the optical afterglow of EP\,240315a is relatively bright compared with this sample, it is comparatively fainter in the X-rays. Still, many GRBs (often not those at the highest redshift) do lie in this region of parameter space. 
Furthermore, the $\gamma$-ray energetics and the location of EP\,240315a on the relations between prompt spectral properties and energetics are also entirely consistent with those seen in GRBs \cite{svinkin24}. 

A striking feature of the spectroscopy is the very sharp break at Ly-$\alpha$ (Figure~\ref{spec}). This is a combination of light from Ly-$\alpha$ in emission and a very low neutral hydrogen gas column density along the line of sight. The lines of sight to many long-GRBs have extremely high column densities of neutral hydrogen gas, exhibiting so-called damped Lyman-$\alpha$ systems with $\log{(N_{\rm H\textsc{i}}/{\rm cm}^{-2})} > 20.3$ \cite{wolfe}. In fact, the median long-GRB has $\log{(N_{\rm H\textsc{i}}/{\rm cm}^{-2})} \sim 21.5$ \cite{tanvir19}. In contrast, the afterglow of EP\,240315a requires a low 
$\log{(N_{\rm H\textsc{i}}/{\rm cm}^{-2})} = 15.9 \pm 0.3$ for $z=4.8585$, although a higher column density is allowed if the absorber's redshift is a free parameter (see Methods). Importantly, this low column 
implies a non-negligible escape fraction of ionising photons. Lyman continuum leakage is not unprecedented in long GRB lines of sight, but very rare \cite{vielfaure20}. At $z \sim 6$ such ionizing photons are required to drive the transition of the intergalactic medium from its neutral to ionised state. However, most studies to date suggest rather low escape fractions from regions of star formation at this cosmic epoch, proving something of a challenge in obtaining the required photon budget. For EP\,240315a the inferred escape fraction may be close to unity (for $\log{(N_{\rm H\textsc{i}}/{\rm cm}^{-2})} = 15.9$). Indeed, because of the relatively bright backlight provided by the counterpart, we can test for this by examining the spectrum at rest-frame wavelengths $<912$\,\AA{} (i.e. the Lyman limit). We detect escaping ionizing photons in the regime 900--912\,\AA{} with a significance of $\sim 4 \sigma$ (see Figure~\ref{leakage} and Methods). This is the highest redshift at which such ionising photons have been directly detected, although the interference of the dense Ly-$\alpha$ forest precludes the robust calculation of the escape fraction from the leaking photons.

\section{Discussion}

EP\,240315a appears very similar to the 
classical cosmological GRBs at comparable redshift. Although it is striking that the EP trigger begins several minutes before the $\gamma$-ray detections, in the absence of this  trigger, the burst would have a typical GRB duration, and similar energetics and would lie on the spectral correlations that hold for long GRBs. A natural conclusion is that this FXT (and by extension a significant fraction of other FXTs) are related to GRBs. Alternatively, if this is not true, and many FXTs are related to other astrophysical phenomena, we should consider if our populations of GRBs are substantially contaminated by different progenitor populations, and if missions such as EP could uncover new classes of high-energy explosions. 

Firstly, we should consider the progenitor of EP\,240315a itself. Although its $\gamma$-ray duration is typical of GRBs, it lasts much longer at softer X-ray energies: 1600~s in the 0.5--4 keV band of the EP WXT. However, it does not have the very long duration at high luminosity $(L_{\rm X}>10^{48}$ erg s$^{-1}$) seen for the candidate relativistic tidal disruption events \citep{burrows11,levan11}. In particular, by the time of the first epoch of \emph{Chandra} observations at 3 days (0.5 days in the rest frame) the luminosity is $L_{\rm X} \sim 10^{46}$ erg s$^{-1}$, two orders of magnitude fainter than \emph{Swift}~J1644+57 at this time. The 1600~s duration would (depending on the classification scheme) place it within the ultra-long GRB population \cite{levan14} which do appear to be related to stellar collapse in some cases \cite{greiner15}. However, the $\gamma$-ray duration is much shorter.

If we identify EP\,240315a as a typical, long-duration GRB then the progenitor would be expected to be a massive star. The presence of Ly-$\alpha$ in emission in the host galaxy suggests that such a young population of stars is present. However, there is good recent evidence which suggests that long-GRBs arise not only from collapsars but also from the merger of compact objects. This is based, in particular on the detection of kilonova signatures in the long duration GRB\,211211A \cite{rastinejad22,troja22} and GRB\,230307A \cite{levan24,yang24}. The appeal of such a progenitor in the case of EP\,240315a would lie in explaining the very low hydrogen column density. In particular, if a progenitor is kicked from its birth site then it may ultimately merge away from the dense dust and gas around the massive stars, presenting a lower hydrogen column density and a higher escape fraction. Thus, the detection of Lyman leakage would be more readily explained at $z \sim 5$ with a merger than with a massive star. Similar constraints may also favour the tidal disruptions of white dwarfs by intermediate-mass black holes, which could take place in (proto)-globular clusters and be relatively gas-free. However, the expected rates of such disruptions are typically even lower than the rates of compact object mergers \cite{mandel22,maguire20}. While these arguments could favour an alternative (non-massive star) progenitor it is important to remember that to reionise the Universe there should be sufficient lines of sight to massive stars with low escape fractions to allow their ionising photons to reach the intergalactic medium. We could, therefore, be observing such a fortuitous line of sight. In principle, the detection of supernova emission could confirm the progenitor, although this would require JWST observations. In the absence of a supernova, the location relative to the host galaxy could be diagnostic. At present our limits on host galaxy emission are weak, at 16 days post-burst, VLT observations provide a magnitude $z=25.44 \pm 0.45$ (AB) for the combination of afterglow and host, implying $M_{{\rm host,}z} > -20.97$. The source location is astrometrically consistent with the afterglow position. However, host detection and offset measurements are within the realm of future observations.  At the current time, aside from noting its observational similarity with long-GRBs, we cannot, with the data available, make strong statements about the progenitor of EP\,204315a. However, we also note that using GRBs as probes of the escape fraction at high redshift will ultimately rely on making robust inferences of progenitors, at least on a statistical level. 

Given our conclusion that EP\,240315a is a normal GRB, it is therefore relevant to consider if many of the observed FXTs could be related to the GRB phenomena. At first sight, there appears to be little in common between the bright GRBs at $z \sim 5$ and the much fainter FXTs. The GRBs have durations of seconds to minutes, are detected predominantly at energies $>10$--100 keV and have peak luminosities of $>10^{52}$ erg s$^{-1}$. The FXTs have durations of minutes to hours, are not detected outside of the soft X-ray regime (probably due to a large delay between the FXT onset and their discoveries) and in the cases where likely redshifts have been measured have luminosities of $L_{\rm X} \sim 10^{46}\mbox{--}10^{47}$ erg s$^{-1}$. However, there are good reasons to believe these populations could be related. A more thorough discussion is provided in the Methods, however, in summary: Current detectors are biased towards GRBs with high peak fluxes, while longer-lived bursts with a lower peak flux, but the same total fluence, are more difficult to detect. Indeed, the luminosity function of GRBs rises steeply to fainter sources \cite{2015MNRAS.447.1911P} so in the narrow, pencil beam surveys undertaken by \emph{Chandra} and \emph{XMM-Newton} we would expect to locate the fainter examples at high redshift where the cosmological volume is maximized. We expect these low luminosity examples to be spectrally soft. While it is not certain that the two populations are related, the presence of a substantial population of on-axis, GRB-like events within the FXT population is entirely plausible (see Figure~\ref{swift_comp}).

Regardless of its progenitor, EP\,240315a points to a new route to identify high-energy transients in the distant Universe. The use of wide-field soft X-ray detectors has long been suggested as a route to enhancing the detectability of high-$z$ GRBs thanks to cosmological redshifting that moves the peak energy of the bursts closer to the X-ray range. EP\,240315a appears to confirm this promise. Although it is only a single event, as the first object with a redshift from EP it has a markedly higher redshift than the median redshifts of \emph{Swift} GRBs ($\bar{z} \sim 2$; \citep{jakobsson06}). EP and similar technologies such as those used in the ESA M7 candidate THESEUS \cite{2021ExA....52..183A} should reveal tens of GRBs beyond $z \sim 6$. If the progenitors of these systems are clear, and in particular if they are massive stars then they offer powerful new routes to probing both the production (e.g. via host galaxies) and absorption of ionizing radiation well into the epoch of reionization. 

Interestingly, narrower field-of-view instruments can also play an important role. \emph{Chandra} has a three-order of magnitude greater sensitivity than the EP WXT, and Athena would be a further factor of $\sim 30$ more sensitive. This substantially offsets the factor of $\sim 8000$ between the field of view of the EP WXT and the Athena Wide Field Imager. The Athena WFI can perform a volume-limited survey for sources with $L_{\rm X} \sim 10^{46}$ erg s$^{-1}$ out to $z >15$ should a burst occur at such a redshift, and scaling from the \emph{Chandra} population allowing for the factor $\sim 4$ in field of view and factor 30 in sensitivity should give a rate of detection of FXTs for Athena that is two orders of magnitude larger than for \emph{Chandra} (although it should be noted that the \emph{Chandra} population has been accrued over 20+ years). Critically, it will also provide precise positions for follow-up with extremely large telescopes. A time domain survey with \textit{Athena} could be a powerful complementary route to discovering the collapse of stars very early in the Universe. 

\section*{Acknowledgements}

Based on observations collected at the European Organisation for Astronomical Research in the Southern Hemisphere under ESO programme 110.24CF (PI Tanvir, Vergani, Malesani);  and with the GTC, under the International Time Programme of the CCI (International Scientific Committee of the Observatorios de Canarias of the IAC), operated on the island of La Palma by the Roque de los Muchachos (program GTC1-23ITP; PIs Jonker, Torres); and with the Italian TNG operated on the island of La Palma by the Fundación Galileo Galilei of the INAF (Istituto Nazionale di Astrofisica) at the Spanish Observatorio del Roque de los Muchachos of the Instituto de Astrofisica de Canarias (program A47TAC\_42, PI Melandri); and with the LBT (program IT-2023B-020, PI Maiorano); and with the Nordic Optical Telescope, owned in collaboration by the University of Turku and Aarhus University, and operated jointly by Aarhus University, the University of Turku and the University of Oslo, representing Denmark, Finland and Norway, the University of Iceland and Stockholm University at the Observatorio del Roque de los Muchachos, La Palma, Spain, of the Instituto de Astrofisica de Canarias. The scientific results reported in this article are based in part on observations made by the \emph{Chandra} X-ray Observatory under program number 25208972 (PI: Levan). We thank the staff of the VLT, GTC, NOT, TNG, LBT and \emph{Chandra} for their critical help in obtaining the data presented here. This work made use of data supplied by the UK \textit{Swift} Science Data Centre at the University of Leicester.

FEB acknowledges support from ANID-Chile BASAL CATA FB210003, FONDECYT Regular 1241005, and Millennium Science Initiative Program  – ICN12\_009.
PDA acknowledges funding from the Italian Space Agency, contract ASI/INAF no. I/004/11/4 and from the GRAWITA INAF Large Program Grant.
AG acknowledges the financial support from the Slovenian Research Agency (P1-0031, I0-0033, J1-8136, J1-2460, N1-0344).
BPG acknowledges support from STFC grant No. ST/Y002253/1.
YDH acknowledges financial support from INAF through the GRAWITA Large Program Grant (ID: 1.05.12.01.04).
LI acknowledges financial support from INAF through the YES Data Grant (ID: 1.05.23.05.15).
PGJ has received funding from the European Research Council (ERC) under the European Union's Horizon 2020 research and innovation programme (Grant agreement No.~101095973).
JK acknowledges funding from the Walter Gyllenberg Foundation of the Royal Physiographical Society in Lund and the Swedish Research Council (VR: Etableringsbidrag 2017-04945).
JM acknowledges financial support of the National Key R\&D Program of China (2023YFE0101200), National Natural Science Foundation of China 12393813, and the Yunnan Revitalization Talent Support Program (YunLing Scholar Project).
DBM is funded by the European Union (ERC, HEAVYMETAL, 101071865). Views and opinions expressed are, however, those of the authors only and do not necessarily reflect those of the European Union or the European Research Council. Neither the European Union nor the granting authority can be held responsible for them. The Cosmic Dawn Center (DAWN) is funded by the Danish National Research Foundation under grant DNRF140.
SPRS gratefully acknowledges support from an STFC PhD studentship and the Faculty of Science and Technology at Lancaster University.
POB and NRT acknowledge support from STFC grant No. ST/W000857/1.
HP acknowledges support by the Spanish Ministry of Science and Innovation with the Ramon y Cajal fellowship number RYC2021-031798-I.
AR acknowledges support from PRIN-MIUR 2017 (grant 20179ZF5KS).
AS acknowledges support from CNES and DIM-ACAV+.
IW is supported by the UKRI Science and Technology Facilities Council (STFC).
VSD and
HiPERCAM are funded by the Science and Technology Facilities Council
(grant ST/Z000033/1).

\begin{figure*}
\begin{center}
\centerline{
\includegraphics[angle=0,width=14cm]{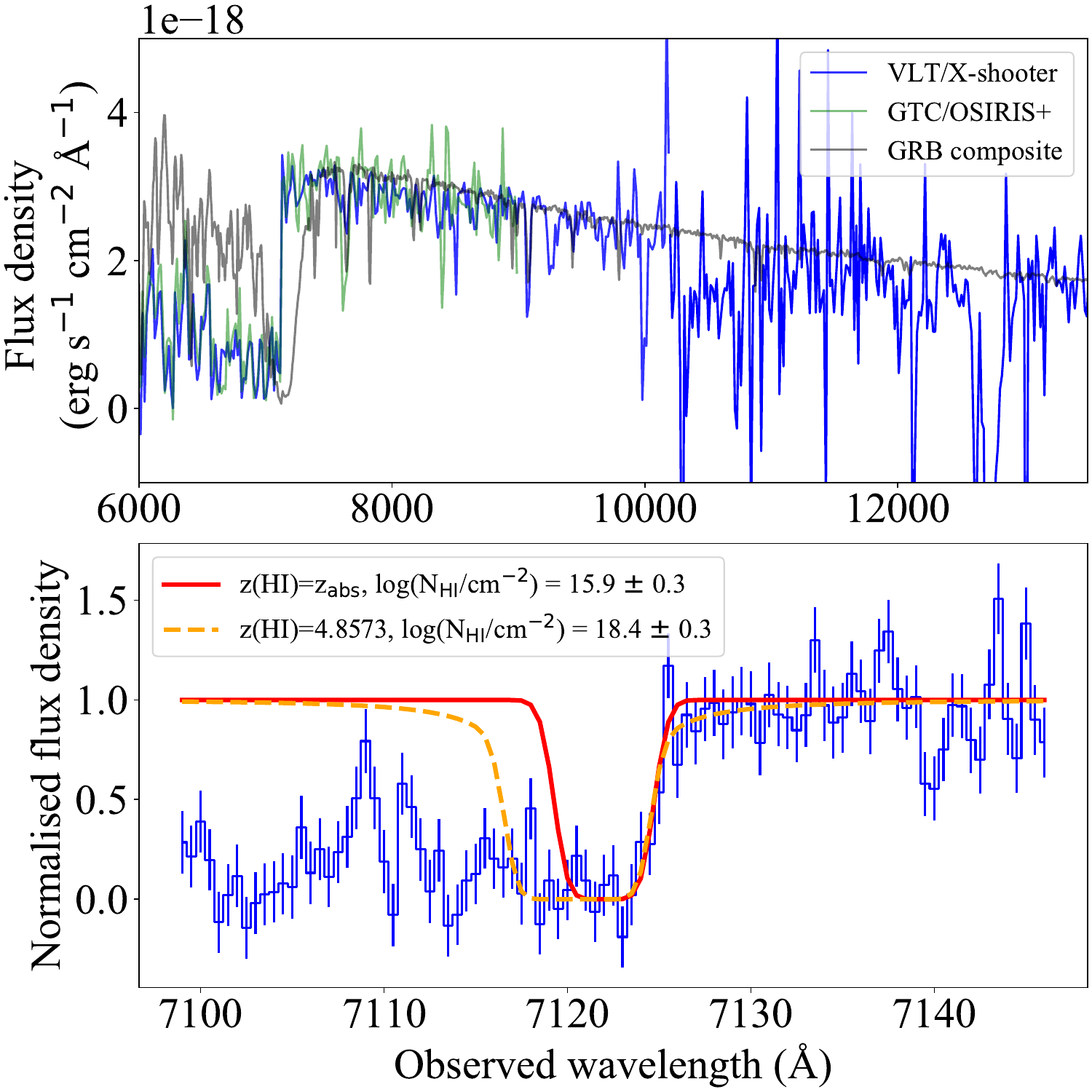}}
\end{center}
\caption{{\bf Optical and IR spectroscopy of EP\,240315a:}. The figure shows GTC/OSIRIS+ and VLT/X-shooter spectroscopy of EP\,240315a obtained 27 and 29 hours after the transient onset, respectively. The strong flux break due to Ly-$\alpha$ is visible at 7120\,\AA. For comparison we over-plot the composite GRB afterglow spectrum of \cite{christensen11}, where we have rescaled the normalised spectrum to have an underlying spectral slope of $F_{\nu} \propto \nu^{-1.0}$. The lower panel shows a zoom-in of the region around Ly-$\alpha$, with the spectrum plotted as the result of the subtraction of our second epoch from the first to remove Ly-$\alpha$ in emission (note this only impacts the spectral shape at Ly-$\alpha$. In addition, we show two fits to the data, one in which the redshift of the neutral hydrogen absorber is fixed to the redshift measured from the metal lines. This requires a very low $\log(N_{\rm H\textsc{i}}/\text{cm}^{-2}) = 15.9 \pm 0.3$. We also plot a model in which the redshift is allowed to vary, in this case we find a solution with a slightly lower redshift and a higher column density of $\log(N_{\rm H\textsc{i}}/\text{cm}^{-2}) = 18.4 \pm 0.3$. The conservative assumption is that we should observe hydrogen and metals at the same location, and the low density and high escape fraction is consistent with the identification of leaking photons blueward of the Lyman limit. }
\label{spec}
\end{figure*}

\begin{figure*}
\begin{center}
\centerline{
\includegraphics[angle=0,width=14cm]{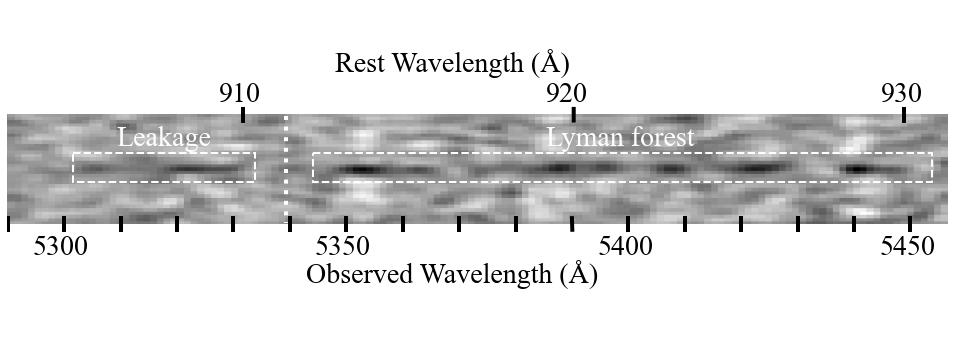}}
\end{center}
\caption{{\bf Lyman continuum leakage from EP\,240315a:} The smoothed 2-dimensional X-shooter spectrum of EP 240315a in the region around the Lyman limit (vertical dotted white line at 912\,\AA{} rest frame). Brighter emission can be seen in the forest, but there is a $4\sigma$ detection of flux from the counterpart blueward of this, consistent with leakage of ionising flux from the host galaxy. }
\label{leakage}
\end{figure*}

    \begin{table*}
    \centering
    \begin{tabular}{ccccc}
    \hline\hline
    Time since FXT & Telescope/instrument & Band & Exposure time & Magnitude \\
    (days) & & & (number $\times$ s) & (AB)  \\
    \hline
1.086 & GTC/OSIRIS+ & $r$ & 6$\times$120 & 21.74 $\pm$ 0.06 \\ 
1.104 & GTC/OSIRIS+ & $i$ & 1$\times$150 & 21.14 $\pm$ 0.04 \\
1.195 & VLT/X-shooter & $r$ & 4$\times$60 & 22.68 $\pm$ 0.05 \\
1.198 & VLT/X-shooter & $g$ & 3$\times$60 & $>$25.0  \\
1.201 & VLT/X-shooter & $z$ & 3$\times$60 & 21.06 $\pm$ 0.06 \\
2.015 & TNG/NICS & $H$ & 42$\times$60 & 21.15 $\pm$ 0.20 \\
2.041 & GTC/HiPERCAM & $g$ & 80$\times$30 & $>25.3$ \\
2.041  & GTC/HiPERCAM & $i$ & 80$\times$30 & 22.32 $\pm$ 0.03 \\
2.041 & GTC/HiPERCAM & $r$ & 80$\times$30 & 23.99 $\pm$ 0.08 \\
2.041 & GTC/HiPERCAM & $z$ & 80$\times$30 & 22.05 $\pm$ 0.03 \\
3.097 & GTC/EMIR & $K_s$ & 672$\times$3 & 21.09 $\pm$ 0.10 \\
3.139 & GTC/EMIR & $J$ & 192$\times$10 & 21.87 $\pm$ 0.15 \\
3.194 & NOT/ALFOSC & $z$ & 27$\times$300 & 22.23 $\pm$ 0.11 \\
4.045 & GTC/EMIR & $K_s$ & 672$\times$3 & 21.47 $\pm$ 0.12 \\
4.052 & TNG/NICS & $H$ & 50$\times$60 & 21.56 $\pm$ 0.19 \\
4.304 & LBT/LUCI & $J$ & 30$\times$63 & 22.49 $\pm$ 0.08 \\
6.414 & LBT/LUCI & $H$ & 48$\times$63 & 23.29 $\pm$ 0.22 \\
7.233 & VLT/FORS2 & $z$ & 32$\times$120 & 23.92 $\pm$ 0.11 \\
16.067 & GTC/OSIRIS+ & $z$ & 45$\times$60 & 24.55 $\pm$ 0.41 \\
16.183 & VLT/FORS2 & $z$ & 22$\times$120 & 25.44 $\pm$ 0.45 \\

    \hline\hline
    \end{tabular}
    \caption{Our optical and IR photometry of EP\,240315a from various ground-based telescopes. See \cite{2021MNRAS.507..350D} for the HiPERCAM filters.}
    \label{tab:photometry}
\end{table*}


\section{Methods}
Below we describe our observations of EP\,240315a with a range of facilities, and also outline in more detail the consistency of the FXT population with low luminosity GRB-like events. Throughout we use the Planck 2018 cosmology \cite{planck18}, magnitudes are given in the AB system and errors at the 1$\sigma$ level unless otherwise stated. 

\subsection{Observations}
EP\,240315a was discovered using the Einstein Probe \citep[EP;][]{EP, Yuan2022} with the first photons detected at 20:10 UT on 15 March 2024 \citep{zhang24}, and announced via the general coordinates network (GCN) approximately 20 hours later. The outburst had a duration of $\approx 1600$\,s as measured by the Wide Field X-ray telescope (WXT) onboard EP. The average flux during this period was $F_{\rm X} = (5.3_{-0.7}^{+1.0}) \times 10^{-10}$ erg s$^{-1}$ cm$^{-2}$ in the 0.5--4 keV band. For a duration of 1600\,s, this implies a fluence of $S_\text{0.5--4 keV} \sim 8.5 \times 10^{-7}$ erg cm$^{-2}$. 

Although there were no coincident $\gamma$-ray triggers reported, a search in data from both the \emph{Swift} Burst Alert Telescope, and the Konus-\textit{Wind} instrument revealed a clear $\gamma$-ray signal which began $\sim 400$~s \emph{after} the onset of activity observed by EP, and lasted for only $\sim 50$~s \citep{svinkin24,delaunay24}. The total burst fluence was $S_\text{20~keV--10 MeV} = (1.63_{-0.40}^{+0.64}) \times 10^{-5}$ erg cm$^{-2}$. 

Serendipitous observations with the ATLAS telescopes 1.1 hours after the onset of the outburst \citep{srivastav24}, discovered a new transient AT\,2024eju, with a cyan filter magnitude of 19.37 $\pm$ 0.14. The probability of uncovering an unrelated bright transient in the small (3 arcminute radius, 90\% containment) error box is low, making this source the prime candidate to be the counterpart of EP\,240315a. 
This identification was confirmed with the detection of the X-ray afterglow by the EP follow-up X-ray Telescope \citep{chen24}, the optical spectroscopy and X-ray observations described below, and the discovery of a radio counterpart \cite{carotenuto24}.

\subsubsection{VLT spectroscopy and photometry}
\begin{figure*}[!ht]
\begin{center}
\centerline{
\includegraphics[angle=0,width=8cm]{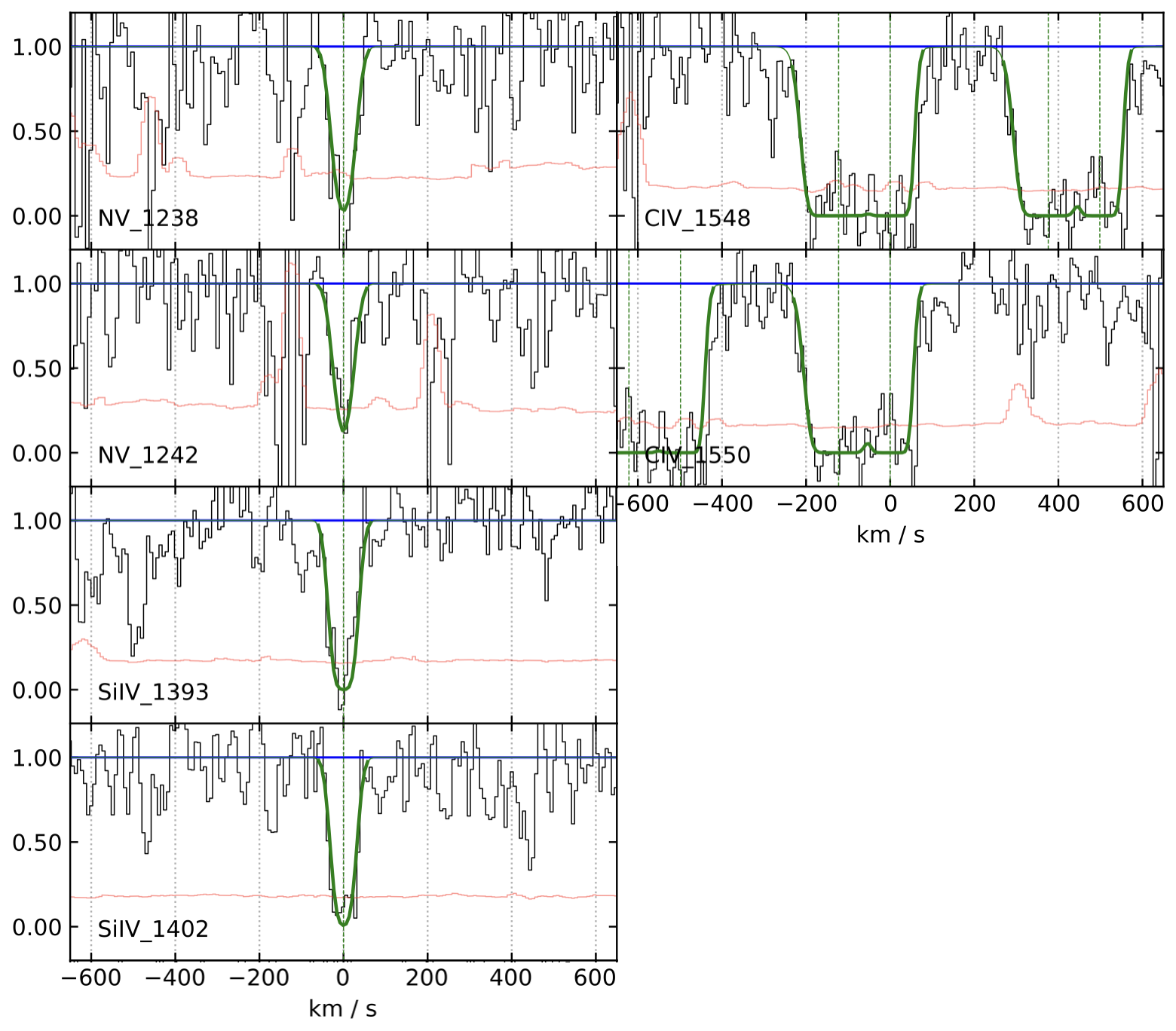}}
\end{center}
\caption{VLT/X-shooter optical spectrum of EP\,240315a at redshift $z=4.859$. All the panels of the high-ionization absorption lines are in velocity space and the zero-velocity has been fixed to $z=4.8585$, corresponding to the single N\,\textsc{v} and Si\,\textsc{iv} transitions. Absorption lines have been fitted with Voigt profiles using the \texttt{Astrocook} software.}
\label{abs_fit}
\end{figure*}

We obtained observations of EP\,240315a with the
ESO VLT and X-shooter \cite{Vernet2011} instrument under programme 110.24CF (PIs Malesani, Tanvir, Vergani), beginning at 01:02:52 UT on 17 March 2024, as soon as the source became visible from Chile. This was 29 hr after the EP source detection, 9 hr after it was reported to the GCN and 3 hr after the ATLAS detection was reported. A faint source was visible in the $r$-band acquisition image, and we obtained spectroscopy of the source with $4\times 1200$ s exposures in the three different arms of X-shooter, covering the wavelength range $3\,000-21\,000$~\AA. The $K$-blocking filter was used to improve the signal-to-noise ratio in the $J$ and $H$ regions of the spectrum. The observations were executed in the ABBA nod-on-slit mode. We adopted the standard X-shooter STARE mode reduction pipeline \citep{Goldoni2006,Modigliani2010} for the UVB, VIS and NIR arms. Then, flux-calibrated spectra for each exposure were adjusted for slit-loss in each arm, with residual sky features subtracted. The individual spectra were then combined into a final science spectrum using the methods outlined in \cite{Selsing19}, with the extraction window (1.6\arcsec) positioned and centered accurately at the location of the trace. Additionally, a telluric correction was applied to the final stacked spectra. All wavelenghts were corrected to the vacuum-heliocentric system. The spectrum is shown in Figure~\ref{spec}, the redshift from quick-look analysis was initially reported in \cite{saccardi24}. The spectrum shows a strong, sharp break at 7120~\AA, which we identify as the Ly-$\alpha$ break at $z=4.859$ refined by several absorption features due to N\,\textsc{v} $\lambda1238$\AA, $\lambda1242$\AA, Si\,\textsc{iv}\,$\lambda1393$\AA, $\lambda1402$\AA{} and C\textsc{iv}\,$\lambda1548$\AA, $\lambda1550$\AA. N\,\textsc{v} and Si\,\textsc{iv} show a single strong component while C\,\textsc{iv} has a larger velocity structure towards the blue (see Figure~\ref{abs_fit}). We fit the identified absorption lines with the \texttt{Astrocook} software \citep{Cupani2020}, a \texttt{Python} code environment to analyze spectra modelled with Voigt profiles, depending on the system redshift $z$, its column density $N$, and its Doppler broadening $b$. The column densities we determined are reported in Table \ref{tab:absorption}.

Other absorption lines from multiple intervening systems along the line of sight have been identified. A strong Mg\,\textsc{ii} foreground systems at $z=2.3050$ (VIS arm) and a C\,\textsc{iv} system at $z=3.6243$ are present. 
Furthermore two other systems are detected at $z=1.9078$ (Mg\,\textsc{ii} and Fe\,\textsc{ii}) and $z=2.5695$ (Fe\,\textsc{ii}).

\begin{table*}[h]
\centering
\begin{tabular}{cc}
\hline\hline
Absorption lines  & $\log(N/{\rm cm}{^{-2}})$\\
\hline
C\textsc{iv}$\lambda1548$, $\lambda1550$$^\dagger$ &$>16.7$ \\
N\,\textsc{v}$\lambda1238$, $\lambda1242$ &$14.6\pm0.2$  \\
Si\textsc{iv}$\lambda1393$, $\lambda1402$ &$14.4\pm0.2$ \\
\hline
\hline
\\
\end{tabular}
\caption{Column densities of high ionization absorption lines identified in the EP\,240315a X-shooter spectrum. They are derived by using a Doppler parameter $b=21\pm2$\,km~s$^{-1}$, obtained from the simultaneous fit of N\,\textsc{v} and Si\,\textsc{iv} single component. $^\dagger$The C\,\textsc{iv} doublet with a different, highly saturated velocity structure shows at least one additional absorption component.}
\label{tab:absorption}
\end{table*}

\begin{figure*}[!ht]
\begin{center}
\centerline{
\includegraphics[angle=0,width=8cm]{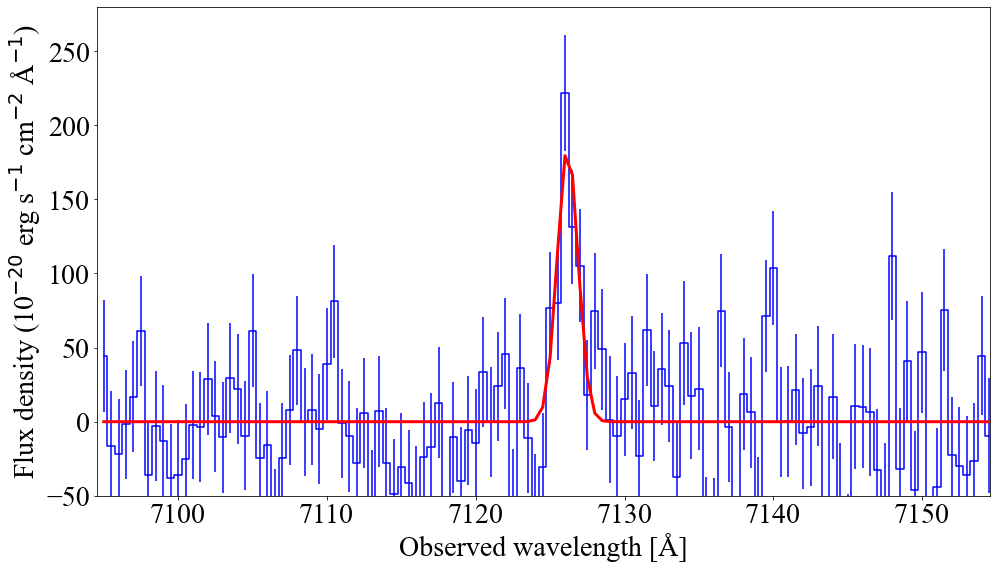}}
\end{center}
\caption{The Lyman-$\alpha$ emission line in the second epoch of X-shooter spectroscopy. The line wavelength of 7126.17 \AA{}  corresponds to a redshift of $z=4.8619$. This likely reflects the bluer emission being absorbed by neutral gas in the host galaxy, however, since the emitting gas may not be at precisely the same velocity (or same location) as the gas observed in absorption, the precise systemic redshift is also uncertain. The detection implies a star formation rate in the host galaxy $>0.75$ $M_{\odot}$ yr$^{-1}$.}
\label{fig:Lyalpha_emission}
\end{figure*}

Aside from the narrow spectral features the continuum is reasonably described by a power-law with $F_{\nu} \propto \nu^{\beta}$ with $\beta=-1.0$. This is typical of the continuum slopes seen in GRB afterglows. 

The break at Ly-$\alpha$ is particularly sharp, suggesting a rather low column density of H\,\textsc{i}. However, it also contains a narrow feature just redward of the Ly-$\alpha$ which suggests a likely contribution from Ly-$\alpha$ in emission in the host galaxy. To remove this contribution we obtained a second epoch of X-shooter observations on 31 March 2024 with an identical set-up (slit position, orientation) to the first epoch. This observation reveals a clear detection of Ly-$\alpha$ in emission (see Figure~\ref{fig:Lyalpha_emission}). The line has a flux of $F =3.2 \times 10^{-18}$ erg cm$^{-2}$ s$^{-1}$, and a narrow width of 1.4~\AA{} (60 km s$^{-1}$). The measured redshift of the line is $z=4.8619 \pm 0.0003$. This suggests that the blue wing is substantially absorbed, and so the actual flux of Ly-$\alpha$ is significantly suppressed by the neutral column, which thus only provides a lower limit on the star formation rate in the host galaxy. This rate is calculated to be $\text{SFR} > 0.43$ $M_{\odot}$ yr$^{-1}$ using the conversion of \cite{vielfaure21}. We note that Ly-$\alpha$ emission originates from the galaxy, and the line of sight is not the same as to the GRB itself. Ly-$\alpha$ could exhibit both higher or lower column densities than the GRB itself, and is best removed from the spectrum prior to $N_{\rm H\textsc{i}}$ fitting. 

Do determine the hydrogen column density on the subtracted spectrum, we model the Ly-$\alpha$ absorption edge using the Voigt-profile fitting code \textsc{VoigtFit} \cite{Krogager2018}. This takes the input spectra and estimates the best-fit column density, $N$, broadening parameter, $b$, and absorption redshift, $z_{\rm abs}$ via a $\chi^2$ minimization approach of the created absorption-line model convolved with the spectral resolution of the data. Fixing the redshift of the Ly-$\alpha$ absorption to that inferred from the more narrow metal lines, $z=4.8585$, yields $\log{(N_{\rm H\textsc{i}}/{\rm cm}^{-2})}= 15.9\pm 0.3$. Leaving the redshift of Ly-$\alpha$ as a free parameter yields a substantially higher column density, $\log{(N_{\rm H\textsc{i}}/{\rm cm}^{-2})} = 18.4\pm 0.3$, mostly due to the preferred lower redshift solution at $z=4.8573$, blueshifted by $\approx 75$\,km\,s$^{-1}$ to the main metal line velocity component. Since the low-ion metal content is expected to be mostly embedded in the neutral hydrogen gas, the low-column-density solution is likely the most appropriate, and would be consistent with the detection of leaking ionising radiation.

A striking feature of the first epoch spectroscopy is that inspection of the UVB arm of X-shooter reveals emission from the transient at wavelengths $<912$\,\AA{} in the rest frame (see Figure~\ref{leakage}). The emission is visible in the range 905--910 \AA{} in the rest frame. To estimate its significance we use an unbinned, non-smoothed stare reduction of the UVB data, and utilize an aperture 10\,\AA{} long and 3 pixels high and compare the measured flux in this aperture to the mean and standard deviation of 20 equally sized apertures placed at locations within $\sim$ 100\, \AA{} of Ly-$\alpha$ but offset from the position of the trace. These results suggest that the leakage is significant at the $>6\sigma$ level. Although the precise significance depends on the choice of central wavelength and aperture size, the overall significance is robust. In the second epoch observations, in which the slit location was identical, we do not observe any sign of emission at this location indicating that any light is arising from the afterglow, and not, for example, from any lower redshift intervening systems.

In addition to our spectroscopy, we also obtained photometric measurements, first with the X-shooter acquisition camera coincident with our first epoch of spectroscopy in the $grz$ filters. We then obtained two further epochs of $z$-band imaging using FORS2 on 23 March and 31 March  2024. FORS2 observations were reduced through the standard ESO pipelines while X-shooter acquisition cameras images were manually corrected for bias and flat-fields. Observations were calibrated against the Pan-STARRS catalogues and results are given in Table~\ref{tab:photometry}.

\subsubsection{Gran Telescopio Canarias (GTC) spectroscopy and photometry }
We obtained observations of EP\,240315a with the OSIRIS+ instrument mounted on the
GranTelescopioCanarias (GTC) under programme GTC1-ITP23 (PIs Jonker, Torres), beginning at 22:55:26 UT on 16 March 2024, about 1 hr after the ATLAS detection of an optical counterpart was reported. Three observations of 1200~s each were obtained using the R1000R volume-phased holographic grating. A slit of 1\arcsec\ width was used. The slit was oriented at the parallactic angle. The data were corrected for bias and flatfield effects and extracted using tasks in {\sc iraf}. Cosmic rays were removed using {\sc lacosmic} \citep{2001PASP..113.1420V}. Wavelength calibration was done using daytime arc-lamp observations using {\sc molly}\footnote{We thank Tom Marsh for the use of 'molly'.}. Flux calibration was done using observations of the spectrophotometric standard star Hiltner~600 taken through a wide slit (2.52\arcsec) on the same night at UT 21:54:58 at similar airmass as the observations of EP\,240315a.  

Photometric observations in the optical were obtained with the GTC using either OSIRIS+ or the HiPERCAM instrument (see Table~\ref{tab:photometry}). In addition, GTC near-infrared (NIR) observations were obtained on two nights using the EMIR instrument on 18 March ($J$ and $K_s$ bands) and 19 March 2024 (only $K_s$). The EMIR images were reduced using scripts based on python and IRAF \citep{Tody1986}. All individual frames are first corrected for readout background in each individual column using overscan information. Flat fields are then created for each observing band and used to correct the scientific data. Flat-corrected science frames are used to create sky background images. The background-only images are normalised to the median flux of each science image and  subtracted from it. Finally, these images are aligned and combined to produce the final frame. The world coordinate system information is obtained using \texttt{astrometry.net} codes \citep{Lang2010}.

\subsubsection{Nordic Optical Telescope (NOT)}

A $z$-band observation was secured using the Nordic Optical Telescope (NOT) at the Roque de los Muchachos observatory (Canary Islands, Spain), using the ALFOSC camera. A standard dithering pattern was used for the observations, and a standard reduction technique was adopted to create the output frame. Photometric calibration was computed by comparison with nearby stars from the Pan-STARRS catalog, with no need to correct for color terms.

\subsubsection{Telescopio Nazionale Galileo (TNG)}
We performed NIR observations of EP\,240315a using the Near Infrared Camera Spectrometer (NICS) camera in imaging mode on the 3.58 m Telescopio Nazionale Galileo (TNG, Canary Islands, Spain) under the long term observation program A47TAC\_42 (PI: A. Melandri). The observation was conducted in two epochs, which started on 2024-3-17 20:10:27 UT (2.0~d after trigger) and on 2024-3-19 20:50:09 UT (4.0~d after trigger), respectively. Hour-long observations in $H$ band were carried out in both epochs. The image reduction was performed using the \texttt{jitter} task of the ESO-eclipse package\footnote{\url{https://www.eso.org/sci/software/eclipse/}}. The photometric measurements using aperture and point-spread function (PSF)-matched photometry were performed using the \texttt{DAOPHOT} package within \texttt{IRAF}. Magnitudes were calibrated with the nearby reference stars in the field of view listed in the Two Micron All Sky Survey (2MASS) catalogue\footnote{\url{https://irsa.ipac.caltech.edu/Missions/2mass.html}}.

\subsubsection{Large Binocular Telescope (LBT)}
We observed the NIR counterpart of EP\,240315a with the Large Binocular Telescope (Mount Graham, Arizona) and the LBT Utility Camera in the Infrared \citep[LUCI,][]{2003SPIE.4841..962S} imager and spectrograph, under program  IT-2023B-020 (PI Maiorano). $J$-band imaging data were obtained on 2024-03-20 at a mid-time 03:30:00 UT, 4.3~d after the EP detection, followed by $H$-band observations at mid-time 2024-03-22T05:20:00, 7.4~d after the trigger. These data were reduced using the data reduction pipeline developed at INAF - Osservatorio Astronomico di Roma \citep{2014A&A...570A..11F} which includes bias subtraction and flat-fielding, bad pixel and cosmic ray masking, astrometric calibration, and coaddition. 

\subsubsection{\textit{Chandra} X-ray Observatory}
We further obtained observations with the \emph{Chandra} X-ray Observatory (DDT programme ID 23495, PI Levan). Two epochs of observations were obtained starting at 19:53 on 18 March 2024 for 9.95 ks and at 04:34 on 26 March 2024 for 18.53 ks. The source was placed at the default location on the ACIS-S3 chip for both observations. 

Data were processed using {\sc ciao} v4.16. Background-subtracted spectra were obtained with the {\sc specextract} script and grouped as 1 count per bin with the {\sc grppha} routine from NASA's High Energy Astrophysics Software \citep[HEASoft;][]{HEASoft} package.
Spectral fitting was performed in {\sc xspec} v12.11.1, using an absorbed power-law model ({\sc tbabs * po}) with Galactic absorption fixed to $N_{\rm H} = 4.44 \times 10^{20}$\,atoms\,cm$^{-2}$ \citep{Willingale13}, abundances from \citet{Wilms00}, and Cash statistics \citep{Cash79}.
For the first epoch, the best fit ($\text{cstat/dof} = \text{26.4/28}$) has a photon index of $\Gamma = 2.0\pm0.5$ and an unabsorbed $0.3$--$10$\,keV flux of $(8.7\pm0.7) \times 10^{-14}$\,erg\,s$^{-1}$\,cm$^{-2}$ as measured with {\sc cflux} (these error bars are at the 67\% confidence level). At the time of the second epoch we only detect a single photon at the source localisation. Using the method of \cite{kraft91} and the measured photon background in the images we obtain a $1\sigma$ range of 0.1--2.2 counts, or a $3\sigma$ upper limit of 7.8 counts. Assuming the same spectrum as observed in the first epoch, this corresponds to a measured flux of $(1.4^{+1.7}_{-1.3}) \times 10^{-15}$ erg s$^{-1}$ cm$^{-2}$.

\subsection{FXT and GRB properties}
At $z=4.859$ the isotropic energy release in the 20 keV -- 10 MeV range is $E_{\rm iso,\gamma} = 7.1 \times 10^{53}$ erg, making the source a relatively energetic GRB. The comparable energy in the X-ray regime (0.5 -- 4 keV, observer frame) is  
$E_{\rm iso,X} = 2.2 \times 10^{53}$ erg, approximately a factor of three lower. Although most high-energy transients detected by \emph{Swift} and \emph{Fermi} today are given GRB designations there has previously been a broader classification regime including GRBs, X-ray rich GRBs (XRRs) and X-ray flashes (XRFs) \citep{heise01}. The distinction between GRBs, XRRs and XRFs was typically drawn at $\log (S_{\rm X} \text{(2--30 keV)}/S_{\gamma} \text{(30--400 keV)}) > -0.5$ \citep[e.g.][]{sakamoto04, 2006A&A...460..653D}, with the bands largely chosen to be convenient for comparing event fluences detected by the \emph{BeppoSAX} WFC and HETE-2. Although we do not have the full spectral information to make these comparisons it seems likely that EP\,240315a does belong to the X-ray rich class of bursts via these diagnostics, despite the relatively high peak energy of $E_p =459_{-155}^{+304}$ keV, although we also note that the rather different time intervals considered by Konus-\textit{Wind} and EP may skew these results. A full analysis of the EP and $\gamma$-ray data could reveal important diagnostics for the GRB prompt emission. 

To put the high-energy emission in the context of \textit{Swift} GRBs, we retrieved from the \textit{Swift} Burst Analyser \cite{Evans2010a} the BAT+XRT light curves of $>480$ GRBs with detected X-ray afterglows and known spectroscopic redshifts. We also downloaded the data products of the high-redshift GRBs 090423, 090429B and 100905A that have only a photometric redshift. In addition, we acquired the X-ray light curves of the relativistic TDEs SwJ1112-82 \cite{Brown2015a} from the Burst Analyser, SwJ1644+57 \cite{Bloom2011a, Burrows2011a,levan11,levan16} from \cite{Mangano2016a}, SwJ2058+05 \cite{Cenko2012a} from \cite{Pasham2015a}, and AT\,2022cmc \cite{Andreoni2022a, Pasham2023a} from \cite{Pasham2023a}. All BAT data products were built using methods from T. Sakamoto and S. D. Barthelmy (NASA/GSFC) and the XRT data with methods from \cite{Evans2007a, Evans2009a}. The density plot was generated using the methods outlined in \cite{Schulze2014a}.

The multi-wavelength properties of EP\,240315a appear to be entirely consistent with those of the GRB population. In particular, the spectral slope from optical/IR photometry $\beta = 1.2 \pm 0.3$ and the X-ray spectral slope of $\beta =1.3 \pm 0.4$ are consistent with each other, although given the large error are also consistent with the cooling break lying between the two bands. However, the overall X-ray to optical spectral slope $\beta_{\rm OX} = 0.9$ at 3 days post transient, which suggests that the X-ray and optical lie on the same branch of the synchrotron spectrum. The optical decay is also entirely consistent with that observed for long GRBs, as is the optical/IR and X-ray luminosity.

\subsection{The link between FXTs and GRBs}
In the main text, we argue that a significant fraction of the FXTs observed previously by \emph{Chandra} and \emph{XMM-Newton} could be low luminosity GRBs typically observed at high redshift (essentially the low luminosity representations of EP\,240315a). This does not mean that the lower-luminosity events must arise from this channel, since the lower photon energies may enable the identification of new or different phenomena. However, we describe below why additional channels (beyond low luminosity GRBs) are not required to explain the observed populations of FXTs with \emph{Chandra} and \emph{XMM-Newton}.

\begin{figure*}[!ht]
\begin{center}
\centerline{
\includegraphics[angle=0,width=10.5cm]{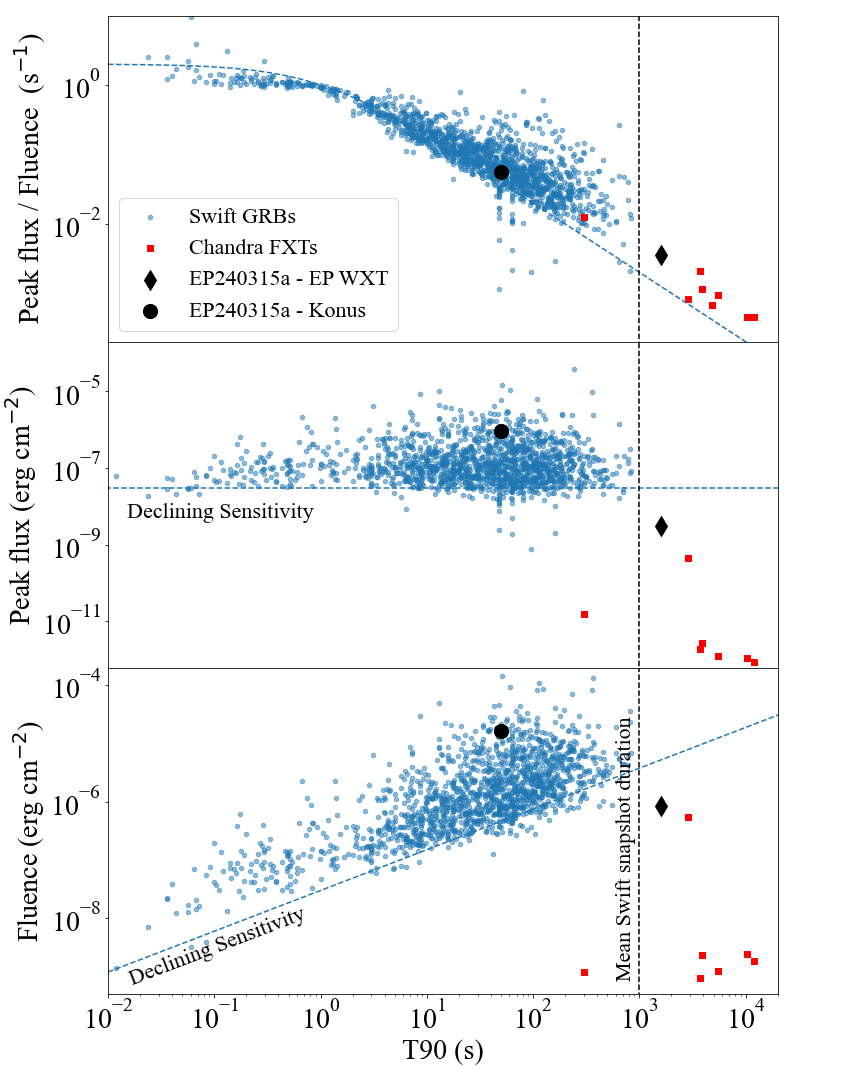}}
\end{center}
\caption{The prompt emission properties of \emph{Swift} GRBs, \textit{Chandra} FXTs \cite{qv22, qv23},  and EP 240315a. The lower panel shows the duration versus the total fluence for \emph{Swift} GRBs demonstrating the strong relationship between $T_{90}$ and total fluence. In particular for very long events, only the highest-fluence (and intrinsically rarest) bursts can be found, providing a bias against typical GRB fluence distributed over longer duration. Furthermore for \emph{Swift}, and also other satellites in low Earth orbit, the time frame over which the integration can occur is limited, with the typical \emph{Swift} observation time marked with the vertical dashed line. The central panel shows the one second peak flux versus duration. Since most triggers are based on the peak flux, this shows little variation with $T_{90}$ as expected. Unsurprisingly given the nature of the detections none of the \emph{Chandra} FXTs fall in the regime where we would have expected detections with \emph{Swift}, even if it had been looking at the location. We note that bursts are plotted in their native discovery bands (15--150 keV for \emph{Swift} and 0.3--10 keV for \textit{Chandra}), but corrections between the two are at most a factor of a few, and not important on this plot. The upper panel shows the ratio of peak flux to fluence versus duration, with the dashed line indicating no evolution for $T_{90} < 1$~s and inverse proportionality afterwards (e.g. for bursts with a given total fluence the peak flux is inversely proportional to the duration). This is unsurprising since for longer lived bursts most energy arises from outside of the main 1s peak. However, interestingly the \emph{Chandra} FXTs as well as EP\,240315a all lie well within the expected bounds of this relationship. }
\label{swift_comp}
\end{figure*}

A striking feature of the FXTs, and indeed EP\,240315a is that they have markedly both longer durations than the GRB populations observed by GRB missions such as \emph{Swift} and \emph{Fermi} ($T_{90}$ is typically $\sim 1000\mbox{--}10000$~s for FXTs compared to 1--100~s for GRBs), and lower peak luminosities, likely in the $L_{\rm } \sim 10^{44} - 10^{47}$ erg~s$^{-1}$ range, although robust redshifts are unavailable for many events. However, these may be selection effects. Firstly, as shown in Figure~\ref{swift_comp}, the EP WXT is substantially more sensitive than the \emph{Swift} BAT for most spectral shapes, and \emph{Chandra} and \emph{XMM-Newton} are more sensitive still. Many normal GRBs detected by the EP WXT will be observable by it for several thousand seconds.

Secondly, most GRB detections to date have been based on rate triggers which favour the identification of high peak fluxes. In principle, the physical mechanisms that create GRBs (e.g.~accretion or magnetar spin-down) are more constrained by the total available energy, rather than the instantaneous flux. Bursts with very long duration but low peak flux are more difficult to detect, and within the \emph{Swift} population there is a strong correlation between duration and fluence, such that very long events can only be detected if they have a very high total fluence, as demonstrated in Figure~\ref{swift_comp}. Indeed, the ultra-long GRB population observed by \emph{Swift} may be under-represented by more than an order of magnitude compared to the true population at the same integrated fluence \cite{levan14}.  This means that ultra-long GRBs are the hard-to-detect tail end of the long-GRB duration distribution \cite{zhang14}. Indeed, the analysis of \cite{zhang14} implies that central engine activity for several thousand seconds is quite common, but that most GRB detectors have higher recovery fractions for bursts with high-peak flux at some point during their outburst. 

The combination of these two effects may naturally create much larger populations of FXTs with durations of thousands of seconds than one would expect by simply extrapolating from the duration distribution of GRBs. 

The differing luminosities of most GRBs and  \emph{Chandra}/\emph{XMM-Newton} population of FXTs can be explained because the latter are found in sensitive narrow field instruments that survey small volumes to extreme depths.  The redshift rate evolution of GRBs detected by \emph{Swift} and \emph{Fermi} is well constrained at brighter luminosities, but despite their far higher source densities, the low-luminosity population is relatively sparsely sampled, and, inevitably is only directly constrained in the local Universe.  \cite{Ghirlanda2022} describe the rate evolution of GRBs across cosmic time ($0.1 < z < 10$), finding a rapid increase in rate as $(1+z)^{3.2}$ out to $z \sim 3$, followed by a comparably rapid decay. This rate density peaks at a higher redshift than the cosmic star formation history, likely because of the metallicity bias in GRB production \cite{fruchter06,kruehler15}. Their population model predicts rate densities for bursts with $L_{\rm X} > 10^{47}$ erg s$^{-1}$. While detection with \emph{Chandra} and \emph{XMM-Newton} is based predominantly on fluence, and not peak flux, the typical peak flux limits used in FXT searches of $F_{\rm X} \sim 10^{-13}$ erg s$^{-1}$ cm$^{-2}$ \cite{qv22,qv23} imply that these surveys are close to volume-limited for the population of events modeled by \cite{Ghirlanda2022} out to $z \sim 10$. 

\begin{figure*}[!ht]
\begin{center}
\centerline{
\includegraphics[angle=0,width=12cm]{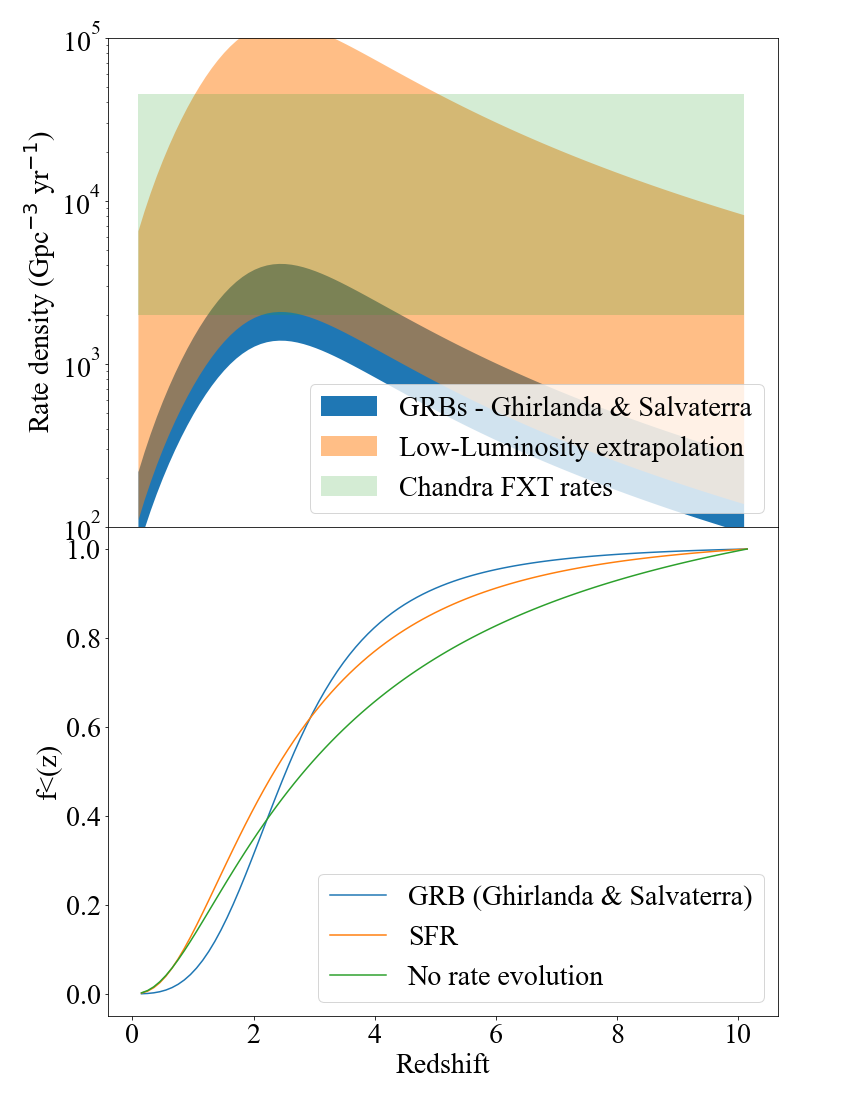}}
\end{center}
\caption{Comparison of GRB and FXT rates. The top panel shows the rate evolution, with the solid blue shows the GRB population model of \cite{Ghirlanda2022}, with the orange indicating its extension to $10^{44}$ erg s$^{-1}$ with slopes between $-1$ and $-1.5$. The solid green box is the rates of FXTs inferred by \cite{qv22}. We note that there are no beaming corrections applied to the GRB rates, so the observed rates are lower by the beaming factor. However, most models predict that the beaming correlates with the energy \cite{Ghirlanda2022}, so we expect lower beaming corrections for low-luminosity GRBs. There are very substantial uncertainties in the rates of all transient types at low luminosity, but low-luminosity GRBs and FXTs are entirely consistent within these. The lower panel shows the expected redshift distribution for a volume-limited survey of events (i.e. assuming all events are recoverable). In this case, we would expect $\sim 50\%$ of events at $z>3$, which would explain the difficulties in locating secure host galaxies in many cases.  }
\label{rates}
\end{figure*}

If we are close to the volume-limited scenario, then the actual number of events recovered in these narrow field surveys is strongly dependent on the behavior of the faint end of the luminosity function. In particular, the GRB luminosity function at the faint end typically has a slope in the range of $-1$ to $-1.5$ \cite{2015MNRAS.447.1911P}, although it is observationally undefined at the faintest levels. \cite{2015MNRAS.447.1911P} provide the rate evolution over the $L \sim 10^{47}-10^{52}$ erg s$^{-1}$ regime. If this were to continue to $10^{44}$ erg s$^{-1}$ the rate increase would be a factor of 1.5--30 (for slopes of $-1$ to $-1.5$, indicating substantial uncertainty due to the unknown slope at these luminosties). Ultimately, an extrapolation of the luminosity function to lower luminosities must decline to avoid over-production of GRB events, for example, in comparison to type Ib/c supernovae, but its extrapolation to the typical peak luminosities seen in FXTs would provide rates (accounting for beaming) of $\sim 10^{3}-10^4$ Gpc$^{-3}$ yr$^{-1}$ at $z \sim 3$, entirely consistent with the uncertain volumetric rates of FXTs observed to date \cite{qv22,qv23}, and substantially lower than the supernova rate. Because of the volume-limited nature of these surveys, we expect a high-redshift population, with a median of $z \sim 3$. In Figure~\ref{rates}, we plot the expected evolution in the rate density for different plausible assumptions including constant rate density, one which follows both the global star formation rate or the low metallicity $Z<0.3 Z_\odot$ and the GRB population model of \cite{Ghirlanda2022}, including an extrapolation to lower luminosity. We also convert these into the expected redshift distributions for volume-limited surveys by assuming that the number of events observed at a given redshift shell is simply $N = V_c R / (1+z)$, where $V_c$ is the comoving volume in the shell and $R$ is the rate density. This demonstrates both the consistency of rates (with substantial uncertainty) and the expectation of a predominantly high redshift population in narrow-deep observations (see Figure~\ref{rates}). Given the difficulty in locating GRB host galaxies at $z>3$, \cite{hjorth12,perley16}, and the typical faintness of the low-luminosity GRB afterglows, the lack of multi-wavelength counterparts and the sparsity of high-confidence host galaxies to many FXTs would be naturally expected in this scenario \cite[e.g.][]{2022MNRAS.514..302E, 2023ApJ...948...91E}.

Since these FXTs are typically faint it will not, in general, be possible to place stringent constraints on the ratio of X-ray to $\gamma$-ray fluxes (see Figure~\ref{swift_comp}, which demonstrates that all lie well below the \emph{Swift} trigger thresholds), and the correlations between spectral peak and isotropic energy and luminosity would imply that these low luminosity events have spectral peaks in the X-ray regime. 
Hence, we conclude that a substantial fraction of the observed extragalactic FXTs can originate in GRB-like events.
In this regard, one can view the  FXTs observed by \emph{Chandra} and \emph{XMM-Newton} as the longer duration tail of the population of XRFs observed by BeppoSAX and WFC, where the differences in duration are largely explained by the selection effects described above.

We note that this is an economical explanation in that it does not require additional progenitor channels beyond those already recognised in the GRB population, and the presence of even lower luminosity GRBs than observed is likely. However, it also does not directly identify any given FXT with a progenitor. Furthermore, it is also plausible that soft X-ray detection, rather than $\gamma$-ray detection, preferentially selects progenitor channels that are until now under-represented in GRB samples.

\end{document}